\documentclass{JHEP3} 
\usepackage{amsmath}
\usepackage{epsfig}
\usepackage{amssymb,amsfonts}

\title{
\boldmath Extended $SL(2,\mathbb{R})/U(1)$ characters, or
 modular properties
of a simple non-rational conformal field theory
\thanks{
Research partially supported by the EEC under the contracts
HPRN-CT-2000-00122, HPRN-CT-2000-00131.}
\unboldmath}
\author{Dan Isra\"el${}^1$, Ari Pakman${}^{1,2}$ and
Jan Troost${}^1$
\\  ${}^1$ Laboratoire de Physique Th\'eorique
de l'\'Ecole Normale Sup\'erieure\thanks{Unit{\'e} mixte  du
CNRS et de l'Ecole Normale Sup{\'e}rieure,
UMR 8549.}  \\ 24, Rue Lhomond  75231
Paris Cedex  05, France\\
and \\
 $ {}^2$ Racah Institute of Physics, The Hebrew University \\
Jerusalem 91904, Israel
$\ $ \\
E-mail:  \email{israel@lpt.ens.fr, pakman@phys.huji.ac.il, troost@lpt.ens.fr }
\\
}

\abstract{We define extended $SL(2,\mathbb{R})/U(1)$ characters which include
a sum over winding sectors. By embedding these characters
into similarly extended characters of $N=2$ algebras,
we show that they have nice modular transformation properties.
We calculate the modular matrices of this simple but non-trivial non-rational
conformal field theory explicitly . As a result, we show that discrete
$SL(2,\mathbb{R})$ representations mix with continuous $SL(2,\mathbb{R})$
representations under modular transformations in the coset conformal field
theory. We comment upon the significance of our results for
a general theory of non-rational conformal field theories.}

\preprint{
LPTENS-04/01\\
RI-01-04
\\hep-th/0402085}

\begin{document}


\section{Introduction}

A systematic theory of rational conformal field theories has been
established. It includes modular transformation properties of
characters, the construction of modular invariant partition functions,
 and, for example, a fairly systematic analysis of boundary conditions consistent with
the chiral algebra. The general theory has proven extremely useful in constructing (supersymmetric)
string theory compactifications~\cite{Gepner:1987qi} and their D-branes (apart from having
applications in two-dimensional physical systems at criticality).

A similarly systematic study of non-rational conformal field theories is
lacking. Recent progress in the $SL(2,\mathbb{C})/SU(2)$ conformal
field theory (see e.g.~\cite{Teschner:1997ft})
as well as in Liouville theory
(see e.g.~\cite{Dorn:1994xn}\cite{Zamolodchikov:1995aa})
and the $SL(2,\mathbb{R})/U(1)$
conformal field theory indicates that such a theory is in
reach. It would seem imperative then to analyze the features of
simple non-rational conformal field theories that have proven
instrumental in the theory of their rational cousins. One
such feature is the modular transformation properties of the
chiral characters of the conformal field theory. We study
these transformation properties for the $SL(2,\mathbb{R})/U(1)$ conformal
field theory in this paper. The $SL(2,\mathbb{R})/U(1)$ conformal
field theory is perhaps the algebraically simplest of non-trivial
non-rational conformal field theories, and it has importance
as providing a black hole background in
string theory~\cite{Elitzur:cb}\cite{Mandal:1991tz}\cite{Witten:1991yr}\cite{Dijkgraaf:1991ba}\cite{Rocek:1991vk}\cite{Giveon:1991sy}.

In a larger framework, it is clear that a better understanding of non-rational
conformal field theories should lead to a firmer grip on non-trivially curved
non-compact  string theory vacua, and therefore to a better understanding
of the stringy theory of quantum gravity. Particular examples directly related to
our paper include cosmological backgrounds and NS5-brane holography.
Let us proceed then to lay bare some of
the basic properties of a simple non-rational conformal field theory to make
a small step in this direction.

We will restrict our computations to the case where the parent
$SL(2,\mathbb{R})$ conformal field theory is at integer level $k$. 
But  our results are extendable at least to the case of generic rational $k$.
Although the WZW $SL(2,\mathbb{R})$ model (and hence the $SL(2,\mathbb{R})/U(1)$ coset)
is well defined for any real $k$,
there are some settings  were the case of integer $k$ is singled out.
In string theory the occurrence of integer level $k$ appears in the context of double scaled little string theory,
where it is the  number of NS5-branes in the string background.
A possible second instance where integer level $k$ values are special
is suggested by the embedding of $SL(2,\mathbb{R})/U(1)$ into non-minimal $N=2$ algebras \cite{Dixon:1989cg}.
The $N=2$ Landau-Ginzburg models
with superpotentials given by the classified 14 singular polynomials of modality one~\cite{Arnold},
flow in the infrared to $N=2$ superconformal field theories with $c={3k \over k-2}$, with integer $k$ \cite{Dixon:1989cg}.
An algebraic description of these models in terms of modular invariants of non-minimal $N=2$ theories
would be a non-rational extension of the well-established equivalence~\cite{Vafa:1988uu} between
$N=2$ LG theories with superpotentials given by the ADE series of singular polynomials with modality 0 and
the ADE modular invariants of minimal $N=2$.

In section~\ref{extended}, we will define the extended characters of the
$SL(2,\mathbb{R})/U(1)$ conformal field theory.
We show in section~\ref{continuous} that
the modular transformation properties of the continuous extended
characters are derived with elementary means. The strategy we
follow for the discrete characters is explained in section~\ref{strategy}.
In section~\ref{decompose} we remind the reader of the connection between the coset
characters and characters for the $N=2$ superconformal algebra. We
will then put to use some known modular properties of the $N=2$
superconformal algebra to derive the modular transformation properties
of the extended discrete $SL(2,\mathbb{R})/U(1)$ coset characters. We thus
construct the modular S- and T-matrices for this simple non-rational
conformal field theory and we derive general lessons from this
example in the concluding section.

\section{The extended characters}
\label{extended}

In this section we define the extended characters for the bosonic
coset $SL(2,\mathbb{R})/U(1)$. A motivation for defining extended characters
is that the regular characters yield a
continuous spectrum of $U(1)$ charges under modular transformations and this problem
is avoided for the extended characters\footnote{The  charges
of the coset are defined with respect to the  $U(1)$ global symmetry that remains after gauging the local $U(1)$.}.
We restrict to discrete and continuous
representations of the parent $SL(2,\mathbb{R})$ conformal field
theory~\cite{Maldacena:2000hw}.
These include all normalizable (i.e. quadratically integrable)
wavefunctions on the group manifold. We introduce the coset central charge:
\begin{eqnarray}
c_{cs} & =& \frac{3k}{k-2}-1 = 2 + \frac{6}{k-2},
\end{eqnarray}
corresponding to a parent $SL(2,\mathbb{R})$ algebra at
level $k$.

\subsection*{Continuous representations}

We label {\it continuous representations} of $SL(2,\mathbb{R})$ by their Casimir
$-j(j-1)$ determined by $j=\frac{1}{2}+is$
(where $s \in \mathbb{R}^+$) and their parity
$\alpha \in {\{} 0, \frac{1}{2} {\}}$.
The characters of the coset can be obtained
by expanding the $SL(2,\mathbb{R})$ characters $\hat{\chi}_{j=1/2+is,\alpha}(q,z)$
into terms with
definite $U(1)$ charge $\alpha + r$ ($r \in \mathbb{Z}$), and
expressing each such term
as a character of a (time-like) $U(1)$ boson
times a character
of the coset\footnote{There are several methods to obtain the coset characters.
The method decribed here is used in~\cite{Pakman:2003kh} (see also~\cite{Ribault:2003ss}).
By embedding the parafermionic algebra into an $N=2$ algebra, a similar
decomposition of the $N=2$ characters into coset and $U(1)$ characters was obtained in~\cite{Bakas:1991fs}.
We make use of this method in Section~\ref{decompose}.
In~\cite{Sfetsos:1991wn} the characters were obtained by counting the
multiplicities of the Hilbert space induced from $SL(2,\mathbb{R})$ and
in~\cite{Griffin:1990fg}
by applying Felder's method to a Feigin-Fuchs representation
of the parafermionic algebra. The coset characters have also been discussed in \cite{Itoh:mt}.}.
This yields
\begin{eqnarray}
\label{carc}
\hat{\chi}_{j=1/2+is,\alpha}(q,z) &=& \mathrm{Tr} \,\,  q^{L_0-(c_{cs}+1)/24} \, z^{J_0^3}  \nonumber \\
 &=& \,q^{s^2 \over k-2} \eta(\tau )^{-3} \sum_{r \in \mathbb{Z}}z^{\alpha+r} \nonumber \\
& = & \sum_{r \in \mathbb{Z}}z^{\alpha+r} \zeta_{r+\alpha}(\tau) \,\, \hat{\lambda}_{1/2+is,\alpha+r} (\tau ) \,,
\end{eqnarray}
with $q=e^{2 \pi i \tau}$ and $\eta(\tau) = q^{1/24} \prod_{n=1}^{\infty} (1-q^{n})$.
The $U(1)$ and the coset characters are:
\begin{eqnarray}
\zeta_{r+\alpha}(\tau) &=& q^{-{(r+\alpha)^2 \over k}} \eta(\tau)^{-1}
\nonumber \\
\hat{\lambda}_{1/2+is,\alpha+r} (\tau ) &=&
\mathrm{Tr} \,\,  q^{L_0 -c_{cs}/24} \nonumber \\
& = &
\eta(\tau )^{-2} q^{\frac{s^2}{k-2}}
q^{\frac{(\alpha+r)^2}{k}} \,.
\end{eqnarray}
It turns out to be natural to define new, extended characters
(in the spirit of \cite{Hwang:1992uk}\cite{Henningson:1991jc}) that
correspond to traces over direct sums of irreducible modules.
 In particular, for  a given continuous representation,
we define the extended characters by summing over all the coset modules whose
 $U(1)$ charges differ by multiples of $k$.\footnote{This
is reminiscent of similar techniques used for
 compact bosons at rational radius squared,
and for characters of the $N=2$ superconformal algebra.}
While the coset characters $\hat{\lambda}_{1/2+is,\alpha+r} (\tau )$ transform
in a complicated fashion under modular transformations,
the  extended  characters have nice
 modular transformation properties, as we will see in the next sections.

We  define the extended characters for the continuous representations as follows:
\begin{eqnarray}
\label{contextchar}
\hat{\Lambda}_{1/2+is,\alpha+r} (\tau ) &=& \sum_{n \in \mathbb{Z}}
\hat{\lambda}_{1/2+is,\alpha+r+kn} (\tau ) \nonumber \\
 &=& \sum_{n  \in \mathbb{Z}}  \eta(\tau )^{-2} q^{\frac{s^2}{k-2}}
q^{\frac{(\alpha+r+kn)^2}{k}} \nonumber \\
&=& \eta(\tau )^{-2} q^{\frac{s^2}{k-2}} \ \Theta_{2(\alpha +r ),k} (\tau,0 )
\end{eqnarray}
where the classical theta functions $\Theta_{m,k} (\tau,\nu)$ at level $k$
are defined as:\footnote{For rational $k$, $k=p/q$, we would introduce
theta functions at level $pq$ and sum over
chiral momentum sectors that differ by $p$ units. This is as in
the case of a chiral boson at rational radius squared.}
\begin{equation}
\Theta_{m,k} (\tau,\nu) = \sum_{p \in \mathbb{Z}+ \frac{m}{2k}} q^{kp^2}
z^{kp}, \ \mathrm{with} \ z=e^{2i\pi\nu}.  \label{defclasstheta}
\end{equation}
Note that for a given $s$, there are $2k$
distinct extended continuous characters (labeled by $\alpha=0,1/2$  and
$r=0,1...k-1$).

\subsection*{Discrete representations}

We consider now  {\it discrete lowest weight representations}\footnote{The discrete representations of $SL(2,\mathbb{R})$ are
actually either highest weight or lowest weight, but both give the
same modules for the coset theory.}
of $SL(2,\mathbb{R})$ labeled by $2j \in
\mathbb{Z}$. Only the representations with $0<j<\frac{k}{2}$ give unitary representations
in the coset theory\footnote{This restriction is further constrained to $\frac12<j<\frac{k-1}{2}$ in the full conformal field
theory \cite{Giveon:1999px, Maldacena:2000hw, Hanany:2002ev}.}.
The $U(1)$ charges are $j+r$ ($r \in
\mathbb{Z}$)
The decomposition of the $SL(2,\mathbb{R})$ discrete character $\chi_{j}(q,z)$ yields (see~\cite{Pakman:2003kh} for details):
\begin{eqnarray}
\label{card}
\chi_{j}(q,z) &=& \mathrm{Tr} \,\,  q^{L_0-(c_{cs}+1)/24} \, z^{J_0^3}  \nonumber \\
 &=& {q^{-\frac{(j-\frac{1}{2})^2}{k-2}} z^{j-\frac{1}{2}} \over \vartheta_1(\tau,\nu)} \nonumber \\
& = & \sum_{r \in \mathbb{Z}} z^{j+r} \zeta_{j+r}(\tau) \,\, \lambda_{j,r} (\tau )
\end{eqnarray}
where\footnote{\label{condition} Both the expression for the discrete $SL(2,\mathbb{R})$
characters in the second line of~(\ref{card}) {\it and}
the decomposition into coset characters in the third line are valid only for $|q|<|z|<1$.}
\begin{equation}
\vartheta_1(q,z)=   z^{-\frac{1}{2}}q^{\frac{3}{24}}
\prod_{n=1}^{\infty}(1-q^n)(1-q^{n-1}z)(1-q^nz^{-1})\,.
\end{equation}
The coset characters are:
\begin{eqnarray}
\label{discrchar}
\lambda_{j, r} (\tau ) &=& \eta(\tau )^{-2} q^{-\frac{(j-\frac{1}{2})^2}{k-2}}
q^{\frac{(j+r)^2}{k}} S_{r} (\tau )\,,
\end{eqnarray}
where the function
$S_r(\tau )$ for integer $r$ is given by:
\begin{eqnarray}
S_r(\tau ) &=& \sum_{s=0}^{\infty} (-1)^s q^{\frac{1}{2} s(s+2r+1)}.
\end{eqnarray}
For the discrete representations we similarly
define extended characters:
\begin{eqnarray}
\Lambda_{j,r} (\tau ) &=& \sum_{n  \in \mathbb{Z}} \lambda_{j,kn+r} (\tau ) \nonumber \\
&=& \eta(\tau )^{-2} q^{-\frac{(j-\frac{1}{2})^2}{k-2}}
\sum_{n  \in \mathbb{Z}} q^{\frac{(j+r+kn)^2}{k}} S_{r+kn} (\tau).
\end{eqnarray}
We now turn to analyzing the modular transformation properties
of these extended characters, which will occupy us for most of
the rest of the paper.

\section{Modular transformation of extended continuous characters}
\label{continuous}
The continuous extended characters in eq. (\ref{contextchar})
are the product of three factors, each having good modular properties.
For T-transformations $(\tau \rightarrow \tau+1)$, we use the formulas
\begin{eqnarray}
\eta(\tau+1) &=& e^{i\pi \over 12} \eta(\tau) \nonumber \\
\Theta_{m,k} (\tau+1,\nu) &=& e^{i \pi m^2 \over 2k }\Theta_{m,k} (\tau,\nu)
\end{eqnarray}
to show that the following T-transformation holds:
\begin{eqnarray}
\hat{\Lambda}_{1/2+is,\alpha+r} (\tau+1) &=&
e^{2 \pi i \left( \frac{s^2}{k-2}-\frac{1}{12} \right)}
e^{2 \pi i \frac{(\alpha+r)^2}{k}}
\hat{\Lambda}_{1/2+is,\alpha+r} ( \tau).
\end{eqnarray}
Under S-transformations $(\tau \rightarrow -1/\tau)$,
using the identities:
\begin{eqnarray}
\eta(-\frac{1}{\tau}) &=& (-i \tau)^{1/2} \eta(\tau) \nonumber \\
e^{-\frac{2 \pi i }{\tau} \frac{ s^2}{k-2}} &=& \sqrt{\frac{-2 i \tau}{k-2}}
\int_{- \infty}^{+\infty}
ds' e^{2 \pi i \tau \frac{{s'}^2}{k-2}} e^{-\frac{4 \pi i s s'}{k-2}}
 \nonumber \\
\Theta_{m,k}  (-\frac{1}{\tau},\frac{\nu}{\tau}) &=&
\sqrt{\frac{-i\tau}{2k}} e^{\frac{\pi i k}{2 \tau} \nu^2}
\sum_{m' \in Z_{2k}} e^{-i \pi m m'/k} \Theta_{m',k} (\tau,\nu).
\label{theta}
\end{eqnarray}
we find that
\begin{eqnarray}
\label{strcon}
\hat{\Lambda}_{1/2+is,\alpha+r} (-\frac{1}{\tau}) &=&
\frac{2}{\sqrt{k(k-2)}} \int_{0}^{+\infty} ds'
\cos (\frac{4 \pi s s'}{k-2}) \nonumber \\
& &
 \sum_{2 \alpha'+ 2r' \in Z_{2k}}  e^{-\frac{\pi i}{k} (2
   r'+2\alpha')(2r+2\alpha)}
\hat{\Lambda}_{1/2+is',\alpha'+r'}(\tau). \label{cosetconttransfo}
\end{eqnarray}
Note that whatever the value of $\alpha$ is, the $2k$ characters with both $\alpha'=0,1/2$
appear in the righthand side of~(\ref{strcon}), with
all the possible values of the continuous parameter $s'$.

We have thus derived the modular transformation properties of the
extended continuous coset characters. It is clear that the continuous
representations transform among themselves under modular
T- and S-transformations. We can thus imagine building modular
invariants
using continuous representations only.

We can check, using the properties of the continuous and discrete
Fourier transform that this part of the S-matrix squares to the
charge conjugation matrix.
Charge conjugation\footnote{Although we haven't explicitly introduced
a variable keeping track of the $U(1)$ charge in our computation,
it can easily be done.} is implemented for
continuous representations by simply changing the sign of the $U(1)$
charge associated to the coset character i.e. $\alpha+r \rightarrow
-(\alpha+r)$.

We now turn to deriving the modular
transformation
properties for the extended discrete characters, which are
less straightforward to compute.
\section{Modular transformations of extended discrete coset
  characters: strategy}
\label{strategy}
For the extended discrete coset characters,
it is still possible to derive the transformation under
the modular T-transformation $\tau \rightarrow \tau +1$ in one step:
\begin{eqnarray}
\Lambda_{j,r} (\tau +1 ) &=&
e^{2 \pi i ( -\frac{(j-1/2)^2}{k-2} - \frac{1}{12})}
e^{2 \pi i \frac{(j+r)^2}{k}}
\Lambda_{j,r} ( \tau ),
\end{eqnarray}
where again we notice diagonality despite the sum over primaries with
differing conformal weight.

To derive the transformation properties of the extended discrete coset
characters under a modular S-transformation $\tau \rightarrow
-\frac{1}{\tau}$, we will follow a more involved strategy. The idea underlying the
computation is to embed the coset model into a model
with a larger symmetry, to derive the transformation properties for
the more symmetrical model, then to extract the modular transformation
properties of the original extended coset characters. Recall  that
the modular properties of the string functions, i.e. of the characters
of the compact and rational $SU(2)/U(1)$ coset, are straightforwardly
derived from the corresponding transformation properties of the
$SU(2)$ parent Wess--Zumino--Witten conformal field
theory (which has a larger symmetry) after decomposing the $SU(2)$
WZW model with respect to a $U(1)$ subgroup.

We choose to embed our model as a sub-sector of an $N=2$ superconformal
field theory. It is well known~\cite{Dixon:1989cg}
that the $SL(2,R)/U(1)$ operators occur
as building blocks for the $N=2$ superconformal algebra with $c>3$.
We will add a free boson to the coset model at the correct radius in
order to enhance the symmetry of the theory to an
$N=2$ superconformal algebra. We then employ the
character decomposition of $N=2$ characters in terms of the compact boson
characters and the bosonic $SL(2,R)/U(1)$
coset characters to link the modules of
$N=2$ to the desired coset modules. We then make use of
known~\cite{Miki:1989ri}~\cite{Eguchi:2003ik}
modular transformation properties of (extended) $N=2$ characters (and of
the compact boson characters) to derive the modular transformation
properties of the constituent discrete bosonic coset characters.

That is a sketch of the strategy we choose, but
one can think of at least two equivalent alternative strategies.
The first would be to embed the coset into the original $SL(2,R)$ theory\footnote{This has been done
implicitly in \cite{Ribault:2003ss}.}, but we would have to be careful with the negative weights
of the time-like $U(1)$.\footnote{Care should also be taken in order not to violate the restriction mentioned in footnote~\ref{condition}.}
The second would be to decompose directly the
lemma we prove in appendix \ref{Miki} in terms of $U(1)$ charges and
apply this directly to the bosonic coset characters.

Our strategy has the advantage
that as a by-product, we will
find useful links between extended $N=2$
characters at $c>3$ and the characters
of our non-compact non-rational conformal field theory (which should
be viewed as analogous to similar Gepner formulas~\cite{Gepner:1987qi}
in the compact, minimal $N=2$ case). In other words, our method lays bare some of the
structure underlying computations in $N=2$ models with $c>3$.

To implement our strategy,
we first turn then to defining the relevant $N=2$ characters in the
next section, then to decompose them in section \ref{decompose},
next to recall their modular transformation properties
in section~\ref{n=2modular}.
Finally, in section~\ref{decomposemodular}
we decompose these modular transformation
properties in terms of the constituent compact boson and coset character.

\boldmath
\section{Embedding $SL(2,\mathbb{R})/U(1)$ in $N=2$ }
\unboldmath
\label{decompose}
The current and primaries of unitary $N=2, c>3$ theories can
by obtained from those of a { \it bosonic} $SL(2,\mathbb{R})$ model by taking out the timelike
boson associated to the bosonization of $J^3$ in $SL(2,\mathbb{R})$, and introducing
a spacelike boson to bosonize the $N=2$ $U(1)$ current.
We refer the reader to ~\cite{Dixon:1989cg} for the details of this construction.
It is shown there that the continuous and discrete  representations of $SL(2,\mathbb{R})$ (with the unitarity bound $0<j<k/2$ in the
discrete),
give rise to all the unitary highest weight representations of the $N=2$
algebra with $c>3$, which were obtained in \cite{Boucher:1986bh} from the Kac determinant.
For a given $j$, there is an infinite number of $N=2$ representations labeled by $r \in \mathbb{Z}$.
The number $r$ is related to the eigenvalue $m$ of $J^3_0$ in $SL(2,\mathbb{R})$
in a way to be indicated below for each representation.

This construction allows to parameterize the
quantum numbers of the $N=2$ representations (the conformal
dimension $h_{j,r}$
and $U(1)$ charge $Q_{j,r}$ of the highest weights)
in terms of $SL(2,\mathbb{R})$ labels,
and this  will be very useful to make explicit the connection between
$N=2$ and $SL(2,\mathbb{R})/U(1)$ characters.
Moreover, as we will show below, the $N=2$ spectral flow (which will play
a central role in our construction) just amounts to a shift in the quantum
number $r$.

The relation between the characters of the $N=2$ representations
and the characters of the coset is similar to the  embedding of the coset into
$SL(2,\mathbb{R})$ considered
in section~\ref{extended}.
Namely, the $N=2$ characters can be expanded into
terms with definite $U(1)$ charge $Q=Q_{j,r} + n$ ($n \in \mathbb{Z}$). Then each term is the product
of a $U(1)$ character (generated by the modes of the R-current $J (z)$) with
highest weight $$\Delta = {3 Q^2 \over 2c}=\frac{k-2}{2k}
\left(Q_{j,r} + n \right)^2,$$ times a coset character.
We will work in the NS sector of the $N=2$ algebra, which suffices for our purposes.

\subsection*{Continuous representations}
From a $SL(2,\mathbb{R})_k$ continuous representation with parameters $j=\frac12 +is,\alpha$ we obtain the $N=2$ representations
\begin{equation}
h_{j,r} = \frac{\frac14 + s^2 + (\alpha + r)^2}{k-2} \,, \qquad Q_{j,r}=Q_m= \frac{2(r+\alpha)}{k-2}
\end{equation}
with $r \in \mathbb{Z}$. The $SL(2,\mathbb{R})$ state which gives rise to each representation has $J^3_0=m=r+\alpha$,
and we will use  $s,m$ to label the representations.
There are no null states
among the descendents \cite{Dobrev:1986hq,Kiritsis:1986rv},
so the $N=2$ character is obtained from the
free action of the modes $L_{-n}, J_{-n}, G^{\pm}_{-n+1/2}$ on the
highest weight state. This yields
\begin{eqnarray}
\label{carcon1}
ch_c(s,m;\tau ,\nu ) &=& \mathrm{Tr}\,\, q^{L_0-c/24}z^{J_0} =
q^{-c/24}  q^{h_{j,m}} z^{Q_{m}} \prod_{n=1}^{\infty}\frac{(1+q^{n-\frac12}z)(1+q^{n-\frac12}z^{-1})}{(1-q^n)^2}
\nonumber \\
&=& q^{-c/24} q^{1/8} q^{h_{j,m}} z^{Q_{m}} \frac{\vartheta_3(\tau,\nu )}{\eta(\tau)^3}
\end{eqnarray}
where
\begin{equation}
\vartheta_3(\tau,\nu ) =
\prod_{n=1}^{\infty}(1-q^n)(1+q^{n-\frac12}z)(1+q^{n-\frac12}z^{-1}) =
\sum_{n=-\infty}^{\infty} q^{n^2 \over 2}z^n \,.
\label{vtheta3}
\end{equation}
Plugging the above identity into the character and using in the exponent of $q$ the identity
\begin{equation}
\label{idcua}
\frac{n^2}{2} + \frac{m^2}{k-2} = \frac{(n-m)^2}{k}
+ \frac{k-2}{2k} \left(\frac{2m}{k-2}  +n \right)^2.
\end{equation}
we find the decomposition of the $N=2$ characters in terms of a
$U(1)$ boson and the coset characters as:
\begin{eqnarray}
\label{carcon2}
ch_c(s,m;\tau ,\nu ) &=&  \eta^{-1}(\tau) \sum_{n  \in \mathbb{Z}} z^{Q_m+n}
q^{\frac{k-2}{2k}(Q_m+n)^2 } \hat{\lambda}_{1/2+is,n-m}(\tau)
\end{eqnarray}
where $\hat{\lambda}_{1/2+is, n-m}$ are the characters of the continuous coset
representation.
We now consider the effect of applying the $N=2$ spectral flow
mapping~\cite{Schwimmer:mf} by $w$ integer units:
\begin{eqnarray}
h & \to & h + wQ + \frac{c}{6} w^2 \nonumber \\
Q & \to & Q + \frac{c}{3} w,
\end{eqnarray}
to both expressions (\ref{carcon1}) and (\ref{carcon2}) of the continuous characters.
We find:
\begin{eqnarray}
q^{\frac{c}{6} w^2} &z^{\frac{c}{3} w}& ch_{c}(s,m;\tau ,\nu+w\tau )
= ch_c(s,m+w;\tau ,\nu )  \nonumber \\
&=& q^{-c/24} q^{1/8} q^{h_{j,m+w}} z^{Q_{m+w}} \frac{\vartheta_3(\tau,\nu )}{\eta(\tau)^3}  \label{flow1} \\
&=& \eta^{-1}(\tau) \sum_{n  \in \mathbb{Z}} z^{Q_m + \frac{wk}{k-2} + n}
q^{\frac{k-2}{2k}(Q_m + \frac{wk}{k-2} +n)^2 } \hat{\lambda}_{1/2+is,n-m}(\tau) \label{flow2} \,,
\end{eqnarray}
respectively, where in the second line we used
$\vartheta_3(\tau,\eta+w\tau)=q^{-\frac{w^2}{2}}z^{-w}\vartheta_3(\tau,\eta)$.
Two observations are in order.
Firstly, we see explicitly that all the representations with fixed
Casimir labeled by $j=1/2 +is$ are mapped into each other under spectral flow.
Secondly, we notice that, in the last expression, eq.~(\ref{flow2}),
summing over spectral flow sectors that differ by $k-2$ units,
corresponds to summing in the bosonic coset characters over sectors that
differ by $k$ units in the chiral momentum.
We are thus led to introduce {\it extended $N=2$ characters} as follows:
\begin{eqnarray}
Ch_c(s,m) &=& \sum_{t\in \mathbb{Z}} ch_c(s,m+t(k-2);\tau ,\nu )  \nonumber \\
&=&
q^{\frac{s^2}{k-2}} \Theta_{2m,k-2}\left( \tau, \frac{2 \nu}{k-2} \right)
\frac{\theta_3(\tau, \nu )}{\eta^{3}(\tau )} \label{context}\\
&=&
\eta^{-1} (\tau ) \sum_{n  \in \mathbb{Z}}
\Theta_{n(k-2)+ 2 m,\frac{k(k-2)}{2}}
\left( \tau, \frac{2 \nu}{k-2} \right)
\hat{\lambda}_{1/2+is, n -m}. \label{decocont} \\
&=&
\eta^{-1} (\tau ) \sum_{n  \in \mathbb{Z}}
z^{Q_m + n} q^{\frac{k-2}{2k}(Q_m + n)^2}
\hat{\Lambda}_{1/2+is,n-m} \label{decocontb}
\end{eqnarray}
We get three equivalent expressions.
Eq. (\ref{context})  is obtained from summing ~(\ref{flow1}) over the spectral flowed sectors.
Eqs. (\ref{decocont})-(\ref{decocontb}) are obtained from two forms of summing~(\ref{flow2}).
These expressions show the intimate connection between
extended $N=2$ characters and extended bosonic coset characters.
Note that the quantum number $2m$ now lies in the set $2m \in \mathbb{Z}_{2(k-2)}$.

\subsection*{Discrete representations}
From a $SL(2,\mathbb{R})_k$ discrete ${\cal D}_j^+$ representation with $0<j<k/2$,
we obtain $N=2$ representations with the following values of
$h_{j,r}$ and $Q_{j,r}$.
\begin{eqnarray}
r \geq 0 && \qquad h_{j,r} = \frac{-j(j-1) + (j+r)^2}{k-2} \qquad \qquad \,\,\,\,\,\,\,\,\, Q_{j,r}= \frac{2(j+r)}{k-2} \label{primdispos} \\
r < 0 && \qquad h_{j,r} = \frac{-j(j-1) + (j+r)^2}{k-2} -r - \frac12 \qquad Q_{j,r}= \frac{2(j+r)}{k-2} -1
\label{primdisneg}
\end{eqnarray}
For $r\geq 0$ ($r<0$), these states are built out of a $J^3$  primary of $SL(2,\mathbb{R})$ with
$J^3_0=m=j+r$ ($J^3_0=m=j+r+1$). Note that the $SL(2,\mathbb{R})$ state with $m=j$ gives
rise to two $N=2$ representations, with $r=0,-1$. These correspond  to
chiral and anti-chiral primaries respectively, as follows from $h_{j,0}=Q_{j,0}/2$ in (\ref{primdispos})
and $h_{j,-1}=-Q_{j,-1}/2$ in (\ref{primdisneg}).

Each representation has one
non-degenerate null descendent at relative $U(1)$ charge
$+1$ ($-1$) for $r\geq 0$ ($r<0$) and level $r+1/2$ ($-r-1/2$)~\cite{Dobrev:1986hq,Kiritsis:1986rv}\footnote{In the cases
$j=0,k/2$ the structure of the descendents is more involved. Modular
properties of the extended $N=2$ character built from the state $j=m=0$, which
is both $N=2$ chiral and anti-chiral, have been described in \cite{Eguchi:2003ik}.}.
For $r=0,-1$, the null states are clearly $G^+_{-1/2}|h_{j,0} \rangle$ and $G^-_{-1/2}|h_{j,-1} \rangle$,
while for other values of $r$ the null states are more complicated (see~\cite{Kiritsis:1986rv} for
explicit expressions for the first levels).

The characters for every $r \in \mathbb{Z}$ are given by
\begin{eqnarray}
ch_d(j,r;\tau ,\nu )
&=& \mathrm{Tr}\,\, q^{L_0-c/24}z^{J_0} \nonumber \\
&=& q^{- \frac{(j-1/2)^2}{k-2} + \frac{(j+r)^2}{k-2}} z^{2(j+r)\over k-2}
\frac{1}{1+zq^{1/2+r}} \frac{\vartheta_3(\tau,\nu )}{\eta(\tau)^3} \,,
\label{cardis1}
 \\
&=& q^{- \frac{(j-1/2)^2}{k-2} + \frac{(j+r)^2}{k-2}-r-\frac12} z^{{2(j+r)\over k-2}-1}
\frac{1}{1+z^{-1} q^{-1/2-r}} \frac{\vartheta_3(\tau,\nu )}{\eta(\tau)^3} \,.
\label{cardis2}
\end{eqnarray}
It is immediate to verify that also for these representations, the spectral flow by $w$ units
is equivalent to a shift $r \rightarrow r+w$.
Both expressions  (\ref{cardis1}) and (\ref{cardis2}) hold for every $r \in \mathbb{Z}$, but each form reflects the structure of the representation
and the null states  for a particular sign of $r$.
For $r \geq 0$ the form (\ref{cardis1}) shows that we have the quantum
numbers (\ref{primdispos}) of the primary state, and the factor
$(1+zq^{1/2+r})^{-1}$ gets rid of the null
state. An analogous statement holds for $r<0$ and the form (\ref{cardis2}).

Another useful expression for the character is:
\begin{equation}
ch_d(j,r;\tau ,\nu )= \frac{\vartheta_3(\tau,\nu )}{\eta(\tau)^3}  \frac{(q^s z)^{\frac{2j-1}{k-2}} q^{\frac{s^2}{k-2}} z^{\frac{2s}{k-2}} }{1+zq^s} \label{useful}
\end{equation}
where $s=r+\frac12$.

Using (\ref{vtheta3}),
we can expand (\ref{cardis1}) into
\begin{equation*}
ch_d(j,r;\tau ,\nu ) = q^{- \frac{(j-1/2)^2}{k-2} +
  \frac{(j+r)^2}{k-2}} z^{\frac{2(j+r)}{k-2}}  \eta(\tau)^{-3} \sum_{n \in \mathbb{Z}}
\sum_{s=0}^{\infty} (-1)^s z^{s + n}  q^{\frac{n^2}{2} +s(\frac{1}{2}+r)}.
\end{equation*}
By shifting $n \rightarrow n-s$ and using (\ref{idcua}) with $m=j+r$, we obtain the decomposition
\begin{equation}
ch_d(j,r;\tau ,\nu ) = \eta^{-1}(\tau) \sum_{n \in \mathbb{Z}}
z^{\frac{2(j+r)}{k-2} +n} q^{\frac{k-2}{2k}\left(\frac{2(j+r)}{k-2} + n\right)^2} \lambda_{j,r-n}(\tau),
\end{equation}
where $\lambda_{j,r-n}(\tau)$
are the characters of the discrete coset representations given in (\ref{discrchar}).

Summing over $k-2$ units of spectral flow, we again obtain three alternative expressions for the discrete
extended $N=2$ characters:
\begin{eqnarray}
Ch_d(j,r) &=& \sum_{n\in \mathbb{Z}} ch_d(j,r+n(k-2);\tau ,\nu )  \nonumber \\
&=&
\sum_{n\in \mathbb{Z}} \frac{q^{- \frac{(j-1/2)^2}{k-2} + \frac{(j+r + n(k-2))^2}{k-2}}
z^{2(j+r+n(k-2)) \over k-2}} {1+zq^{1/2+r+n(k-2)}} \frac{\vartheta_3(\tau,\nu )}{\eta(\tau)^3}  \label{discext}\\
&=&
\eta^{-1} (\tau ) \sum_{n  \in \mathbb{Z}}
\Theta_{n(k-2)+ 2(j+r),\frac{k(k-2)}{2}}
\left( \tau, \frac{2 \nu}{k-2} \right)
\lambda_{j, r-n} (\tau ) \label{decodisc} \\
&=&
\eta^{-1} (\tau ) \sum_{n  \in \mathbb{Z}}
z^{\frac{2(j+r)}{k-2} + n} q^{\frac{k-2}{2k}\left(\frac{2(j+r)}{k-2} + n\right)^2}
\Lambda_{j,r-n} (\tau ).
\end{eqnarray}
and we have chosen the fundamental domain $r \geq 0, r \in
\mathbb{Z}_{k-2}$. 
The form (\ref{discext}) can be easily recast into (see (\ref{useful}))
\begin{eqnarray}
Ch_d(j,r;\tau ,\nu )&=&
\frac{\vartheta_3(\tau,\nu )}{\eta(\tau)^3} \sum_{s \in r + \frac12 + \mathbb{Z}(k-2)}
\frac{(q^s z)^{\frac{2j-1}{k-2}} q^{\frac{s^2}{k-2}} z^{\frac{2s}{k-2}} }{1+zq^s} \nonumber \\
= \frac{\vartheta_3(\tau,\nu )}{\eta(\tau)^3} &&
\frac{1}{k-2} \sum_{r' \in \mathbb{Z}_{k-2}} e^{-2\pi i \frac{ r'
    (r+1) }{k-2}} I \left(\frac{4}{k-2},\frac{r'}{k-2} -\frac12,
\frac{2j-1}{k-2} -\frac12; \tau,\nu  \right), \nonumber\\
\end{eqnarray}
where the function $I(k,a,b;\tau,\nu)$ is defined in eq. (\ref{funcI}).

\subsection*{Relation to \cite{Eguchi:2003ik}}
In this subsection, we show the relation between our extended
$N=2$ characters with those introduced in~\cite{Eguchi:2003ik}.
We first recall that in~\cite{Eguchi:2003ik},
super-Liouville theory was studied as an $N=2$ superconformal
theory with central charge $c= 3 \hat{c} = 3 + 3 {\cal Q}^2$
with ${\cal Q} = \sqrt{2K/N}$ where $K$ and $N$ are
positive integers (with greatest common divisor equal to
one)\footnote{Only in this subsection  will
$N$ denote this positive number and does not refer to the number of
supercurrents in the superconformal algebra.}.
To link that study to ours, we identify the
central charges of the respective $N=2$ algebras. For integer level
$k$, this leads to\footnote{The $N=2$ techniques of~\cite{Eguchi:2003ik}
  can
also be used to treat the case of rational values of $k$.}:
\begin{eqnarray}
c &=& \frac{3k}{k-2} = 3 + \frac{6K}{N} \nonumber \\
K &= & 1 \nonumber \\
N & = & k-2.  \label{link}
\end{eqnarray}
The characters themselves are matched by identifying the quantum numbers as follows.
Our $U(1)$ charge $Q_{j,r}$ of the highest weight is $Q$ in~\cite{Eguchi:2003ik}.
For the continuous characters  (\ref{context})  our quantum numbers $\frac{s^2}{k-2}$ and $2m \in Z_{2(k-2)}$
correspond to  $\frac{p^2}{2}$ and $j \in Z_{2NK}$ in eq.(2.25)  of~\cite{Eguchi:2003ik} (''massive characters'').
For the discrete characters, our quantum numbers $(j,r)$ correspond to $(s/2,r)$ in~\cite{Eguchi:2003ik}.
The identity between our discrete extended character~(\ref{discext}) and
eq.(2.19)  of~\cite{Eguchi:2003ik}
(''massless characters''), can be seen by rewriting~(\ref{discext}) as
\begin{equation}
Ch_d(j,r )
=
 \sum_{n\in \mathbb{Z}} \frac{\left\{z q^{(k-2)\left( n+
    \frac{2r+1}{2(k-2)} \right)}\right\}^{
\frac{2j-1}{k-2}} }{1+z q^{(k-2)(n+ \frac{2r+1}{2(k-2)})} }
z^{2 \left( n+ \frac{2r+1}{2(k-2)} \right)}
    q^{(k-2) \left( n + \frac{2r+1}{2(k-2)} \right)^2}
\frac{\theta_3(\tau ,\nu )}{ \eta^{3}(\tau)}.\\
\label{extdiscr}
\end{equation}
Several useful properties of the extended characters are listed in~\cite{Eguchi:2003ik}.
We now turn to the next step in the program, which is to rederive
their modular transformation properties. Finally, we will
use the decomposition formulas to dissect the modular transformation rules.

\section{The $N=2$ modular transformations}
\label{n=2modular}
The modular transformation S of the continuous extended $N=2$
characters can be derived using the same techniques as we employed for the
extended continuous coset characters. We define the formula for
the (radial) momentum
$s$ in terms of the conformal dimension and $U(1)$ R-charge:
$s^2 /(k-2) = h' - [(2m)^2+1/4](k-2),$ where
the $U(1)$ charge is: $Q=2m/(k-2)= 2(\alpha +r )/(k-2)$. We then find~\cite{Eguchi:2003ik}:
\begin{eqnarray}
Ch_c(s,m ; -\frac{1}{\tau}, \frac{\nu }{\tau}) &=&
\frac{2}{k-2} \ e^{i \pi \frac{k}{k-2} \frac{\nu^2}{\tau}  }
\sum_{2m' \in Z_{2 (k-2)}} e^{- \frac{ 4i\pi }{k-2} m m'} \nonumber \\
& & \int_0^{\infty} ds' \cos \left( \frac{4 \pi s s'}{k-2} \right)
Ch_c(s',m'; \tau,\nu ). \label{n=2conttransfo}
\end{eqnarray}
To derive the modular transformation properties of the discrete
extended $N=2$ characters, we  will need a useful lemma given
in appendix~\ref{Miki}.\footnote{
A proof of this lemma is given in~\cite{Miki:1989ri}
for square tori and real $\nu$. We extend the proof
to the generic case in appendix~\ref{Miki}. Note that this provides a
firm backing of the results in e.g. \cite{Eguchi:2003ik} \cite{Ahn:2003tt}.}
Using this lemma it is possible to rederive the
modular transformation properties as in~\cite{Eguchi:2003ik}.
We only use the restricted case of relevance to us in the following.
The extended discrete $N=2$ characters transform into discrete and
continuous characters as follows\footnote{In the computation, the principal value prescription is
implemented at the point where we split the momentum integral into
negative and positive momenta. In recombining the two integrals,
we have eliminated a possible pole contribution (which
the principal value prescription in the appendix~\ref{Miki}
tells us to do) at $s'=0$. The
final expression can be shown to be regular at $s'=0$.}
\footnote{
Note that the ``partial range'' of quantum numbers appearing on the right
hand side, discussed in~\cite{Eguchi:2003ik}, namely $1 \le 2j' \le k-1$,
coincides in the particular case under study with the full
range~\cite{Hanany:2002ev} of
extended discrete characters necessary to describe all discrete
representations
in the coset.
It is no surprise that all relevant characters appear
on the right hand side, as all characters
in the unitary supersymmetric coset model will satisfy the lower bound on
the conformal dimension associated to the partial range discussed
in~\cite{Eguchi:2003ik}. It is nevertheless striking that the
precise bound on the discrete representation label is naturally
reproduced from an analysis of generic discrete $N=2$ characters.}  :
\begin{eqnarray}
e^{-i \pi \frac{c \nu^2}{3 \tau} } && Ch_d(j,r;-\frac{1}{\tau},\frac{\nu}{\tau})
=
 \frac{1}{k-2} \sum_{2m' \in Z_{2 (k-2)}} e^{- \frac{ 2i\pi
    }{k-2} (2j+2r)m'}
\int_0^{\infty} ds'
\nonumber \\ & &
\frac{\cosh \pi \left( s' \frac{k-4j}{k-2} + im' \right) }{
\cosh \pi (s'+ i m')}
 Ch_c(s',m';\tau,\nu)
\nonumber \\
                              & & +
\frac{i}{k-2} \sum_{r' \in Z_{k-2}} \sum_{2j'=2}^{k-2}
e^{- 2 \pi i \frac{(2j+2r)(2j'+2r')-(2j-1)(2j'-1)}{2(k-2)}}
Ch_d(j',r'; \tau, \nu)
\nonumber \\
                              & & +
\frac{i}{2(k-2)} \sum_{r' \in Z_{k-2}}
e^{- 2 \pi i \frac{(2j+2r)(2r'+1)}{2(k-2)}} \times \nonumber \\
&& \ \ \times \ \left[ Ch_d(1/2,r';\tau,\nu)-Ch_d((k-1)/2,r';\tau,\nu) \right]
\label{n=2discrtransfo}
\end{eqnarray}
We thus recalled the method for the derivation of the
modular transformation properties of the extended $N=2$ characters,
and have strengthened the backbone of the proof by extending the
lemma in  appendix~\ref{Miki} to generic modular parameters.
The modular transformation properties have also been clarified slightly,
by labeling the characters by natural $SL(2,R)/U(1)$ labels.
It is appropriate here to recall that the extended $N=2$ characters
were introduced in~\cite{Eguchi:2003ik} to have a discrete spectrum
of R-charges in the right-hand side of the modular transformation
of discrete characters, eq.~(\ref{n=2discrtransfo}).

\subsection*{Check for $N=2$}
A useful check on the above modular transformation property for
discrete characters is given by computing whether
the square of the modular S-matrix is the charge conjugation matrix.
We can perform this check by (artificially) distinguishing a
discrete and a continuous sector, and by viewing the S-matrix as
being upper triangular:
\begin{eqnarray}
S &=& \left( \begin{array}{cc} A & B \\ 0 & D \end{array} \right),
\end{eqnarray}
where $A$ indicates the coefficients of the discrete characters
contributing
to the modular transform of discrete characters, etcetera.
The technical details are recorded in appendix~\ref{ssquared}.
We find, after using the character identities:
\begin{eqnarray}
Ch_c(s=0,r+1/2;\tau,\nu) &=&
Ch_d(1/2,r;\tau,\nu)+Ch_d((k-1)/2,r;\tau,\nu)
\label{n=2split}\,,  \\
Ch_d(j,r;\tau,\nu) &=& Ch_d (k/2-j,-r-1; \tau,-\nu)\,, \nonumber \\
Ch_c (s,m; \tau,\nu) &=& Ch_c (s,-m;\tau,-\nu)\,,
\end{eqnarray}
that the matrix $S^2$ is indeed the charge conjugation matrix,
which changes the sign of the $U(1)$-charge. In the appendix
we show in detail how the split into continuous and discrete
representations indeed turns out to be artificial,  which can be
expected on the basis of the splitting of the continuous $N=2$
representation with $s=0$ and odd $U(1)$ R-charge into two discrete
ones (see equation (\ref{n=2split})). This
is a faithful foreshadowing of what will happen in the computation
for the coset.
\section{The modular transformations decomposed}
\label{decomposemodular}
We now wish to plug the decomposition rule for the $N=2$ characters
recalled in section \ref{decompose} 
in terms of coset characters into the modular transformation rules
which we rederived in section \ref{n=2modular}.
\subsection*{Continuous transformations revisited}
We want to start with the simple case of rederiving the modular
transformation rules for extended continuous coset characters
using $N=2$ characters. 
The analysis of the modular transformation properties of the
continuous characters of the coset goes as follows:
\begin{eqnarray}
Ch_c \left( s,\alpha+r; -\frac{1}{\tau},\frac{\nu}{\tau} \right)
&=& \frac{1}{\sqrt{-i \tau} \eta(\tau)} \sum_n
\sqrt{\frac{-i \tau}{ k (k-2)}} e^{ \pi i \frac{k}{k-2} \frac{\nu^2}{\tau}}
\sum_{m \in Z_{k(k-2)}} \nonumber \\
& & e^{-2 i\pi  \frac{m (n(k-2)+2 (\alpha+r))}{k(k-2)}}
\Theta_{m,\frac{k(k-2)}{2}} \left( \tau, \frac{2 \nu}{k-2} \right) \nonumber \\
& &
\hat{\lambda}_{1/2+is,
n-(\alpha+r)} (-1/\tau ) \nonumber \\
 & = &  \frac{1}{\eta(\tau)} e^{i \pi \frac{k \nu^2}{ \tau(k-2)}}
\frac{2}{k-2}
\sum_{q \in Z_{2(k-2)}} e^{-2 \pi i q (\alpha+r)/(k-2)}
\nonumber \\
&& \int_0^{\infty}  ds' \cos \left(\frac{4 \pi s s'}{k-2} \right)
\nonumber\\
&& \sum_n \Theta_{n(k-2)+q,\frac{k(k-2)}{2}} \left(\tau, \frac{2 \nu}{k-2} \right)
\hat{\lambda}_{1/2+is',n-q/2} (\tau )\nonumber \\
\end{eqnarray}
where the first line follows from the use of the modular
transformation properties of $\Theta$-functions (\ref{theta}),
 and the second line by
using the transformation property of the $N=2$ characters
(\ref{n=2conttransfo}) (which we
decomposed into coset characters using (\ref{decocont})).
We should identify coefficients of truly linearly independent
$\Theta$-functions (of which there are $k(k-2)$ at level $k(k-2)/2$).
To that end, we need to split the sums in the right hand side as
$n'=k \tilde{n} + p $ where $n \in \mathbb{Z}$ and
$p \in \mathbb{Z}_k$, while we split $q = (k-2) q'+t$, where $q' \in \mathbb{Z}_2$ and
$t \in \mathbb{Z}_{k-2}$. We then shift $p+q' \rightarrow
p$ and find:
\begin{eqnarray}
\sum_n e^{- 2 \pi i m n/k}
\hat{\lambda}_{1/2+is, n- (\alpha+r)}(\tilde{q})
 =
2 \sqrt{\frac{k}{k-2}}
\sum_{\tilde{n}} \sum_{q' \in \mathbb{Z}_{2}} \int_0^{\infty}  ds'
\cos(\frac{4 \pi s s'}{k-2})
 \nonumber \\
e^{-2 \pi i t/k (\alpha+r)} e^{4 \pi i p (\alpha+r)/k}
e^{- 2 \pi i q'(\alpha+r)}
 \hat{\lambda}_{1/2+is',k \tilde{n}+p-k q'/2-t/2} (q).
\end{eqnarray}
where we have identified: $m\equiv t$ mod $k-2$ and
$m -t \equiv  (k-2)p$  where $p \in \mathbb{Z}_k$.
We can read the obtained result as a modular
transformation property of continuous coset characters. After performing a
discrete Fourier transform on $m$ (by summing over $m\in \mathbb{Z}_k$),
we recognize on the left hand side the extended continuous characters we
defined; on the right hand side the remaining constrained sum over
$kq'+t-2p$ gives the sum over $2(\alpha'+r') \in \mathbb{Z}_{2k}$.
We realize that this formula agrees exactly with the
result~(\ref{cosetconttransfo}) we derived in another manner previously.
Thus we have rederived the transformation property of extended continuous
characters of the $SL(2,R)/U(1)$ coset
in an admittedly roundabout fashion, but we have learned useful
techniques to be applied in the case of the extended discrete characters.
\subsection*{Modular transformations of discrete extended coset characters}
Now we have made all the necessary preparations to calculate the
modular transformation properties of extended discrete bosonic coset
characters. We follow the same reasoning as we just
did
for the continuous representations. By applying on the one hand the
modular transformation properties of $\Theta$-functions (\ref{theta}),
 and on the
other hand the modular transformation properties of $N=2$ characters
(\ref{n=2discrtransfo}),
we obtain:
\begin{eqnarray}
e^{- \pi i \frac{k}{k-2} \nu^2/\tau} Ch_d (j,r; -\frac{1}{\tau},\frac{\nu}{\tau})
&=& \frac{1}{ \eta(\tau)} \frac{1}{ \sqrt{k (k-2)}} \sum_{n \in \mathbb{Z}}
\sum_{m \in \mathbb{Z}_{k(k-2)}} \nonumber \\
& & e^{-2 i\pi  \frac{m (n(k-2)+2 (j+r))}{k(k-2)}}
\Theta_{m,\frac{k(k-2)}{2}} \left( \tau, \frac{2 \nu}{k-2} \right) 
\hat{\lambda}_{j,-n+r} (-1/\tau ) \nonumber \\
 & = &  \frac{1}{\eta(\tau)} 
\frac{1}{k-2}
\sum_{q \in \mathbb{Z}_{2(k-2)}} e^{-2 \pi i q (j+r)/(k-2)} \nonumber \\
& &
\int_0^{\infty}  ds' \ \frac{\cosh \pi \left( s' \frac{k-4j}{k-2}+i
  \frac{q}{2} \right)
}{\cosh \pi (s'+i \frac{q}{2})} 
\nonumber \\
& & 
 \sum_{n' \in \mathbb{Z}} \Theta_{n'(k-2)+q,\frac{k(k-2)}{2}} 
\left( \tau, \frac{2 \nu}{k-2} \right)
\hat{\lambda}_{\frac{1}{2}+is',n'-\frac{q}{2}} (\tau )
\nonumber \\
& + & \frac{i}{k-2} \frac{1}{\eta (\tau )} \sum_{r'\in Z_{k-2}} \sum_{2j'=2}^{k-2} 
e^{- \frac{ \pi i}{k-2} [(2j+2r)(2j'+2r')-(2j-1)(2j'-1)]}
\nonumber \\
& & 
\sum_{n' \in \mathbb{Z}} \Theta_{n'(k-2)+2j'+2r', \frac{k(k-2)}{2}} 
\left( \tau,\frac{2 \nu}{k-2} \right) 
\lambda_{j',-n'+r'} (\tau )
\nonumber \\
& + &  \frac{i}{2(k-2)} \frac{1}{\eta (\tau )} \sum_{r'\in Z_{k-2}}  
e^{- \frac{ \pi i}{k-2} (2j+2r)(2r'+1)}
\nonumber \\
 & & \left\{ 
\sum_{n' \in \mathbb{Z}} 
\Theta_{n'(k-2)+1+2r', \frac{k(k-2)}{2}} \left( \tau, \frac{2
\nu}{k-2}\right) \lambda_{\frac{1}{2},-n'+r'} (\tau ) - \right.
\nonumber \\
 & &  \left. \sum_{n'\in \mathbb{Z}} 
\Theta_{n'(k-2)+k-1+2r', \frac{k(k-2)}{2}} \left( \tau, \frac{2
\nu}{k-2}\right) \lambda_{\frac{k-1}{2},-n'+r'} (\tau) \right\}   
\label{step}. \nonumber\\
\end{eqnarray}
As in the continuous case, we
equate the two alternative forms of the modular transform. We then
fix $m$ in the first expression and equate coefficients of
linearly independent $\Theta$-functions. The identification of
linearly independent contributions is tedious but straightforward.
As in the case of continuous characters we then perform a discrete
Fourier transform and find:
\begin{eqnarray}
\Lambda_{j,r} (-1/\tau )
 &=& \frac{1}{\sqrt{k(k-2)}} \sum_{2(\alpha'+r') \in \mathbb{Z}_{2k}}
e^{\frac{4 \pi i}{k} (j+r)(\alpha'+r')} 
\int_0^{\infty}  ds' \frac{\cosh \pi \left(s' \frac{k-4j}{k-2}+i
  \alpha' \right)
}{\cosh \pi(s'+i \alpha')}
\hat{\Lambda}_{\frac{1}{2}+is',\alpha'+r'} \nonumber \\
& + & \frac{i}{\sqrt{k(k-2)}} \sum_{2j'=2}^{k-2} \sum_{r' \in Z_{k}}
e^{ -\frac{4 \pi i}{k} (j+r) (j'+r')} e^{ \frac{4 \pi i}{k-2}
  (j-1/2)(j'-1/2)}
\Lambda_{j',r'} (\tau )\nonumber \\
& + & \frac{i}{2 \sqrt{k(k-2)}}  \sum_{r' \in \mathbb{Z}_{k}}
\left\{ e^{ -\frac{4 \pi i}{k} (j+r) (\frac{1}{2}+r')}
\Lambda_{\frac{1}{2},r'} (\tau )
- e^{ -\frac{4 \pi i}{k} (j+r) (-\frac{1}{2}+r')}
\Lambda_{\frac{k-1}{2},r'} (\tau ) \right\} \nonumber\\
\end{eqnarray}
This is the quantity we wanted to calculate, namely the modular
transform of the extended discrete character. The most important
and new qualitative aspect of this formula is the mixing
between discrete and continuous representations.
The method we used can be extended to at least all rational rational
values of $k$, albeit with some technical refinements.
\subsection*{Check for coset}
The modular S-matrix is again upper-block diagonal, when we list the
representations as a column vector $(discrete, continuous)$. When
we name the blocks in the S-matrix $A,B$ and $D$ for the
left-upper, right-upper and right-lower block, we have
already checked earlier (in section \ref{continuous})
that the matrix $D$ squares to the charge conjugation
matrix in the continuous sector.
The calculation for $S^2$ is done in detail in the appendix.
Using the character identity
\begin{eqnarray}
\hat{\lambda}_{\frac12,\frac12+r}  &=& \lambda_{\frac{k-1}{2},-r} + \lambda_{\frac12,-r-1}
\end{eqnarray}
one can then show that the charge conjugation matrix
 exchanges $(j,r)$ with $(\frac{k}{2}-j,-r)$
for all $j$ in the discrete sector. Again, the split of the
continuous representation with odd parity into two discrete
 representations
turned out to be crucial in allowing for a contribution from the
 ``continuous'' sector to the discrete sector, thus making for the
expected charge conjugation matrix.\footnote{Note
that we have the character identity:
\begin{eqnarray*}
\lambda_{j,r} &=& \lambda_{k/2-j,-r},
\end{eqnarray*}
which shows that charge conjugate representations have equal
characters,
as required.}

\section{Conclusions}
The main result of our paper is the formula for the modular
transformation matrix $S$ for both extended continuous and extended
discrete $SL(2,R)/U(1)$ coset characters. We were able to derive it
by embedding the $SL(2,R)/U(1)$ modules into modules of an enlarged
$N=2$ superconformal algebra. An interesting aspect
of the modular transformation matrix is that, given the presence
of discrete representations, it inevitably links those up with both
discrete and continuous representations. This is a qualitatively new
and important property of the characters of these representations.

It seems important to generalize these results to the case of generic
rational and irrational values of the level $k$, and possibly, to
the case of non-extended coset characters. For the latter , one could apply
our method on the transformation rules for ordinary $N=2$ characters
(see e.g. \cite{Ahn:2003tt}).

As we have argued, we also view these computations as non-rational
analogues of the derivation of the modular transformation properties
of string functions, and their connection to the $N=2$ characters
with $c<3$. We have thus laid bare a
non-trivial modular S-matrix of a non-rational
conformal field theory, and its connection to $N=2$ characters with
$c>3$. We believe that the features of this $S$-matrix that we
uncovered have analogues in many other examples of more complicated
non-rational conformal field theories.

At the heart of our paper lies an extension of the computation
of \cite{Miki:1989ri} which is based on the  technique
of shifting an integration contour over momenta and picking up
poles corresponding to discrete representations on the way over.
This is highly reminiscent of techniques used in other analyses of
non-rational conformal field theories (see
e.g.
\cite{Maldacena:2000kv}\cite{Hanany:2002ev}\cite{Israel:2003ry}\cite{Ribault:2003ss}).

Given our modular S-transformation matrix one can now wonder whether
an analogue of the Verlinde-formula would hold for this non-rational
conformal field theory and whether one can classify modular invariants
(in particular in the discrete sector). One can also use the
modular S-matrix to study possible boundary states \cite{Ribault:2003ss}
(even) more systematically (as has been done for the rational case).
It would indeed be interesting to study the
systematics of boundary conformal field theories that can be
constructed in non-rational conformal field theories
once the modular S-matrix is known.

Finally, our study should be a useful tool
to analyze the conjectured \cite{Giveon:1999px} / proven
\cite{Hori:2001ax}\cite{Tong:2003ik} duality between the
supersymmetric coset and $N=2$ Liouville.


\section*{Acknowledgments}
Thanks to Costas Bachas, Shmuel Elitzur, Hershel Farkas, David Kazhdan, Elias Kiritsis,
Boris Pioline, Sylvain Ribault, Volker Schomerus, and especially Costas Kounnas for
comments and interesting discussions.
AP is supported in part by the Horowitz Foundation. This work is
supported in part by the Israeli Science Foundation.

\appendix
\boldmath
\section{Lemma for $N=2$ transformations}
\unboldmath

\label{Miki}
We follow \cite{Miki:1989ri} and derive the modular
transformation properties of the following function:
\begin{eqnarray}
I(k,a,b;\tau,\nu ) &=& \sum_{r \in Z + \frac{1}{2}}
e^{2 \pi i (a +\frac{1}{2})(r+\frac{1}{2})}
\frac{(zq^r)^{b+\frac{1}{2}}}{1+z q^r} z^{kr/2}
q^{kr^2/4}  \label{funcI}
\end{eqnarray}
where $a,k \in \mathbb{R}$, $k>0$  and $b \in \mathbb{C}$.
We parameterize $\nu$ as: $\nu = \nu_1 -\tau \nu_2$,
$\nu_{1,2} \in \mathbb{R}$. Our aim is to extend the
proof of \cite{Miki:1989ri} \--~given for $\tau \in i\mathbb{R}$ and
$\nu \in \mathbb{R}$~\-- to generic complex values.
We would like to rewrite the S transform of this function as
a contour integral in the following way:
\begin{eqnarray}
B&=&
\frac{i}{\tau} e^{-\frac{ik\pi\nu^2}{2 \tau}}
I\left( k,a,b; -\frac{1}{\tau},
\frac{\nu}{\tau} \right)
\nonumber \\
&=& \frac{e^{i\pi a}}{2 i \pi} \left[ \int_{\mathcal{C}_{-\epsilon}}
+ \int_{\mathcal{C}_{+\epsilon}} \right] dZ \
\frac{\pi}{2} \frac{e^{- 2\pi b Z
- 2i\pi a (i\tau Z - \nu )
+ \frac{2i\pi \tau k}{4} Z^2}}
{\cosh \pi Z \ \cos \pi (i\tau Z-\nu )}
\label{contourint}
\end{eqnarray}
\begin{figure}[!h]
\begin{center}
\epsfig{figure=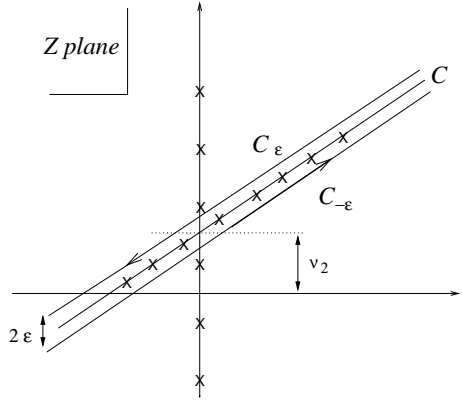, width=60mm}
\caption{Choice of contour of integration (for $\tau_1>0$).}
\label{cont}
\end{center}
\end{figure}
where the parameter $\epsilon$ is infinitesimal and positive.
The contours encircles the line $\mathcal{C}:\ Z =i y / \tau+i\nu_2$,
$y \in \mathbb{R}$ (Fig.~\ref{cont}).
This equality is valid because the contour is such that we pick up all
the poles of the integrand inside the contour of integration.
These are the zeroes of $\cos \pi (i\tau Z -\nu )$ which
occur at $i\tau Z -\nu = -y - \nu_1 \in \mathbb{Z} + 1/2$.
At these poles, labeled by a
half-integer $r$: $$Z = \frac{i}{\tau} (r- \nu )$$
we find residues
$$(-1)^{r+\frac{1}{2}} \frac{i}{\tau}
\frac{e^{2\pi i a (r+1/2)- 2 i\pi b (r-\nu)/\tau} e^{- ik\pi
    (r-\nu)^2/(2\tau )}  }{ e^{i\pi (r -\nu )/\tau} + e^{-i\pi
      (r-\nu )/\tau}},$$
which leads to the equality quoted above. This equality is valid
as long as there are no poles coming from the $\cosh \pi Z$ factor
on the contour of integration. These special cases occur only
for $\nu_2 \in \mathbb{Z}+1/2$.
We will not consider these special cases, which occur in the
Ramond sector and are not crucial for our purposes.
We now note that we have the
expansions:
\begin{eqnarray}
\frac{1}{2 \cos \pi(i\tau Z - \nu )} &=& \sum_{n=0}^{\infty} (-1)^n
e^{\pm (2n+1) \pi i (i\tau Z - \nu )}
\end{eqnarray}
whenever $| e^{\pm i 2 \pi (i\tau Z - \nu )} |$ is smaller than $1$.
This is true on the contours $\mathcal{C}_{\mp \epsilon}.$
We then plug these expansions into the right hand side of
eq.~(\ref{contourint}) and find:
\begin{eqnarray}
B &=& \sum_{r \in Z+ a +\frac{1}{2}} \frac{1}{2 \pi i}
\int_{\mathcal{C}} dZ \ J^b(r;Z)
\label{BJ}
\end{eqnarray}
where $$J^b(r;Z) = i \pi e^{i \pi r} z^{r}
\frac{e^{-2\pi b Z+2\pi \tau r Z+2i\pi \tau\frac{kZ^2}{4}}}{
\cosh(\pi Z)},$$ since:
\begin{eqnarray}
B &=& \frac{1}{2 \pi i}  \int_{\mathcal{C}_{+\epsilon}}
 dZ \pi \sum_{n=0}^{\infty}
(-1)^n e^{-(2n+1) \pi i (i\tau Z- \nu)}\frac{e^{\pi i a - 2 \pi b Z -
2 \pi i a(i\tau Z -\nu )+ 2i\pi \tau\frac{kZ^2}{4}}}{\cosh \pi Z}
\nonumber \\
& &
+ \frac{1}{2 \pi i}  \int_{\mathcal{C}_{-\epsilon}}
 dZ \pi \sum_{n=0}^{\infty}
(-1)^n e^{+(2n+1) \pi i (i\tau Z- \nu )}\frac{e^{\pi i a - 2 \pi b Z -
2 \pi i a(i\tau Z -\nu )+ 2i\pi \tau\frac{kZ^2}{4}}}{\cosh \pi Z}
\end{eqnarray}
and we can identify $r=n+a +\frac{1}{2}$
and $r=-n+a -\frac{1}{2}$ in the first and
second term respectively to obtain formula~(\ref{BJ}).
When taking $\epsilon \rightarrow 0$
we also add a minus sign to the contour $\mathcal{C}_{+\epsilon}$
and switch its direction.

Now, in order to obtain the expression for the
characters in the right hand side,
we wish to {\it (i) } shift the $\mathcal{C}$ contour of integration
by $\frac{2 i r}{k}-i\nu_2$,
\begin{figure}[!h]
\begin{center}
\epsfig{figure=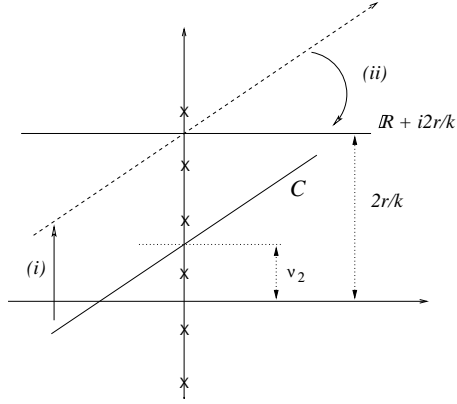, width=60mm}
\caption{Change of contour of integration (case $r>\frac{k \nu_2}{2}$).}
\label{defcont}
\end{center}
\end{figure}
for each term indexed by $r$ and {\it (ii)} tilt it parallel
to the real axis (Fig.~\ref{defcont}).
We will then eventually obtain the desired quadratic term $q^{r^2 /k}$
which will match onto the quadratic term of continuous characters.
The poles we pick up on this occasion arise from the
zeroes of the factor $\cosh \pi Z$, at values of $Z$ such that
$Z=is$ with $s \in \mathbb{Z}+ \frac{1}{2}$. When a
pole falls exactly on the shifted contour we will count it  with half 
its value and define the tilted integral over the real axis  as a principal value
(effectively accounting for the other half of the pole). We
obtain two terms for $B$: one term, $J_1^{a,b}$, corresponds to the
contribution of the poles.
The second term, $J_2^{a,b}$,
corresponds to the shifted contour.
This last term is easily obtained:
\begin{eqnarray}
J_2^{a,b} &=& \sum_{r \in Z+a +\frac{1}{2}} \frac{1}{2 \pi i}
P \int_{\mathcal{C}+\frac{2ri}{k}-i\nu_2}
dZ \ J^b(r;Z) \nonumber \\
&=& \sum_{r \in Z+a +\frac{1}{2}} e^{i \pi r} z^r q^{r^2/k} P
\int_{\mathbb{R}} dx\
\frac{e^{- 2 \pi (b +\frac{1}{2}) (x+ 2r i/k)}
q^{k x^2/4}}{1+ e^{-2 \pi(x+2ri/k)}}
\end{eqnarray}
where in the last line the contour has been tilted to obtain an integral
along the real axis. No poles are crossed during
this deformation. Thus we obtain an integral over continuous
characters of $N=2$. \\

In the term $J_1^{a,b}$, the poles contribute positively
for $r >\frac{k \nu_2}{2}$, and negatively for $r <\frac{k \nu_2}{2}$:
\begin{eqnarray}
J_1^{a,b} &=& (\sum_{r>\frac{k \nu_2}{2}} \sum_{2r/k \geqslant s >\nu_2}-\sum_{r<\frac{k \nu_2}{2}}
\sum_{2r/k \leqslant s <\nu_2}) \epsilon(s-\frac{2r}{k}) j^b(r,s)
\end{eqnarray}
where $j^b(r,s)$ is the residue of $J^b(r,Z)$ at the pole $Z=is$
(where $s \in \mathbb{Z} +\frac{1}{2}$)
and $\epsilon(\gamma)=\frac{1}{2}$ for $\gamma=0$ and
$\epsilon(\gamma)=1$ otherwise (which takes into account the principal value
prescription).

In this expression for $J_1^{a,b}$, we wish to
perform the sum over $r$ for
fixed $s$.
Let us begin with the case $r>\frac{k \nu_2}{2}$.
For every  $s >\nu_2$ we need to determine those
$r>\frac{k \nu_2}{2}$, $r \in Z + a+\frac{1}{2}$ such that
$ s \leqslant 2r/k $
(we assume $\nu_2 >0$, but the opposite case is treated in
the same way).
There is always a number  $\delta(a,s) \in [-\frac12, \frac12) $
such that the pole in $s$ is picked for the values of $r$
given by
$r=R+ks/2+\delta(a,s)$ where $R = \frac12, \frac32, \ldots$

Then we consider the case $r<\frac{k\nu_2}{2}$. For every $s<\nu_2$, we have similarly
to sum over $r=R+\frac{ks}{2} + \delta(a,s)$ (where  now $\delta(a,s)  \in (-\frac12, \frac12]$  ), and
$R = -\frac12, -\frac32, \ldots$

We thus obtain that $J_1$ becomes (with $r,s$ half-integer):
\begin{eqnarray}
J_1^{a,b} &=& \left( \sum_{s>\nu_2} \sum_{R>0}  -
\sum_{s< \nu_2} \sum_{R<0} \right)
\epsilon(R + \delta(a,s ))
j^b(ks/2 + R + \delta(a,s ),s) \nonumber \\
 &=& \left( \sum_{s>\nu_2} \sum_{R>0}  -
\sum_{s<\nu_2} \sum_{R<0} \right)
\epsilon(R + \delta(a,s ))
(-i) e^{2 \pi i (\frac{k}{4}-b+\frac{1}{2})s}
 (e^{i \pi} z q^s)^{R+\delta(a,s)} z^{\frac{ks}{2}}
q^{\frac{k s^2}{4}}
\nonumber \\
\end{eqnarray}
Noting that $|zq^s|=e^{-2\pi \tau_2(s-\nu_2)}$ and
using the expansions:
$$1/(1+x) =\sum_{n=0}^{\infty} (-x)^n
\ \mathrm{for} \ |x|<1$$
$$1/(1+x) =-\sum_{n=1}^{\infty} (-x)^{-n}
\ \mathrm{for} \ |x|>1$$
we can sum on $R$. For the values of $s$ such that
 $|\delta(a,s)| \neq \frac{1}{2}$ we get:
\begin{eqnarray}
J_{1 \alpha}^{a,b} &=& \sum_{\substack{s \in \mathbb{Z}+\frac{1}{2} \\ |\delta(a,s)| \neq \frac{1}{2}}}
e^{2 \pi i (k/4-b+\frac{1}{2})s}
e^{i \pi \delta(a,s)}
\frac{(z q^s)^{\frac{1}{2}+\delta(a,s)}}
{1 + z q^s}
z^{ks/2} q^{ks^2/4}
\end{eqnarray}
while for both cases $\delta(a,s )=\pm \frac{1}{2}$, we get
\begin{eqnarray}
J_{1 \beta}^{a,b} &=& \frac{1}{2} \sum_{\substack{s \in \mathbb{Z}+\frac{1}{2} \\ |\delta(a,s)| = \frac{1}{2}}}
e^{2 \pi i (k/4-b+\frac{1}{2})s}
e^{- i \pi/2} \frac{1-z q^s}{1+ z q^s} z^{ks/2} q^{ks^2/4}.
\end{eqnarray}
so that
\begin{equation}
J_{1 }^{a,b}= J_{1 \alpha}^{a,b} + J_{1 \beta}^{a,b}
\end{equation}
This is the lemma of \cite{Miki:1989ri} extended to generic $\tau$ and $\nu$.

For comparison with \cite{Eguchi:2003ik}, note that naming  the function used
in~\cite{Eguchi:2003ik} to obtain the modular transformation of the
discrete extended characters $I^{ES}(k,a,b;{\tau},{\nu})$, we have
\begin{eqnarray}
I^{ES}(k,a,b;{\tau},{\nu})
&=&  e^{-i \pi a} I (2k,a-\frac{1}{2},b-\frac{1}{2};{\tau},{\nu}) \nonumber \,\,.
\end{eqnarray}

\boldmath
\section{Computing $S^2=C$}
\unboldmath
\label{ssquared}
\boldmath
\subsection*{Modular S-matrix of $N=2$}
\unboldmath
The details of the computations of the square of the S-matrices
is given in this appendix. For the computation of $S^2$ for the
$N=2$ extended characters,
we simplify the computation slightly, and note on beforehand
that the factors $e^{i \pi c/3 \nu^2/\tau} \times e^{i \pi c/3
  (\nu/\tau)^2 (-\tau)}$ cancel out in $S^2$, such that we don't
need to take them along. Next, we distinguish the following
block matrices in the upper-triangular modular matrix $S$:
\begin{eqnarray}
{A_{j,r}}^{j',r'} & =& \frac{i}{k-2} e^{ -\frac{\pi i }{k-2}
  ((2j+2r)(2j'+2r')-(2j-1)(2j'-1))} \quad \mathrm{for} \quad 2j'=2,3,...,k-2
\nonumber \\
&=&  \frac{i}{2(k-2)} e^{ -\frac{\pi i }{k-2}
  (2j+2r)(1+2r')} \quad \mathrm{for} \quad 2j'=1
\nonumber \\
&=&  -\frac{i}{2(k-2)} e^{ -\frac{\pi i }{k-2}
  (2j+2r)(1+2r')} \quad \mathrm{for} \quad 2j'=k-1
\nonumber \\
{B_{j,r}}^{s',2m'} & =&  \frac{1}{k-2} e^{ -\frac{2i\pi}{k-2}
  (2j+2r)m'} \frac{\cosh \pi \left(s'+i m' + \frac{2s' (1-2j)}{k-2}\right)
}{\cosh \pi (s'+im')}
\nonumber \\
{D_{s,2m}}^{s',2m'} & =&   \frac{2}{k-2} e^{ -\frac{4i\pi}{k-2}
  mm'} \cos \frac{4\pi s s'}{k-2}.
\end{eqnarray}
We can then compute the squares of $A$ and $D$:
\begin{eqnarray}
{A^2_{j,r}}^{j'',r''} &=& \delta_{j,k/2-j''} \delta_{r,-r''-1}
 \quad \mathrm{for} \quad 2j''=2,3,...,k-2
\nonumber \\
                    &=& \frac{1}{2} \delta_{j,(k-1)/2}
 \delta_{r,-r''-1} - \frac{1}{2} \delta_{1/2,j} \delta_{r,-r''-1}
\quad \mathrm{for} \quad 2j''=1
\nonumber \\
&=& \frac{1}{2} \delta_{1/2,j}
 \delta_{r,-r''-1} - \frac{1}{2} \delta_{(k-1)/2,j} \delta_{r,-r''-1}
\quad \mathrm{for} \quad 2j''=k-1
\nonumber \\
{D^2_{s,2m}}^{s'',2m''} & =&  \delta_{2m+2m''}^{k-2} \delta(s-s'').
\end{eqnarray}
Next, we calculate the off-diagonal block in the square
of the modular matrix $S$:
\begin{eqnarray}
{(B \cdot D)_{j,r}}^{s'',2m''} &=& \frac{1}{2(k-2)} \delta_{2j+2r+2m''}^{k-2}
[p] \left\{ \sin^{-1} \frac{\pi}{k-2} (2j-1+2is'') \right.
\nonumber \\
&+& \sin^{-1} \frac{\pi}{k-2} (2j-1-2is'')
 +(-1)^p \left( \cot \frac{\pi}{k-2} (2j-1+2is'') \right.  \nonumber\\
&&  \qquad \qquad \left. \left. +\cot
\frac{\pi}{k-2} (2j-1-2is'' )\right)\right\}  \quad \mathrm{for} \quad 2j=2,3,...,k-2
\nonumber \\
& =& \frac{1}{4} \sum_{\alpha'}   \delta_{1+2r+2m''}^{k-2}
[p] (-1)^{2 \alpha' p} \delta^+ (s'') \quad \mathrm{for} \quad 2j=1
\nonumber \\
&=& \frac{1}{4}  \sum_{\alpha'}   \delta_{1+2r+2m''}^{k-2}
[p] (-1)^{2 \alpha' (p+1)} \delta^+(s'') \quad \mathrm{for} \quad 2j=k-1
\end{eqnarray}
where $\delta^+$ denotes the delta-function on the positive
real half-line. We have also introduced the Kronecker symbol mod $k$;
$\delta^{k}_{m} [p]$ means that $m = kp$.
We also find:
\begin{eqnarray}
{(A \cdot B)_{j,r}}^{s'',2m''}&& = \frac{1}{2(k-2)} \delta_{2j+2r+2m''}^{k-2}
[p] \left\{ -\sin^{-1} \frac{\pi}{k-2} (2j-1+2is'') \right.    \nonumber \\
& & \left.
-\sin^{-1}
\frac{\pi}{k-2} (2j-1-2is'') \right.    \nonumber \\
& & \left.
 \!\!\!+ (-1)^p \left( -\cot \frac{\pi}{k-2} (2j-1+2is'')
- \cot \frac{\pi}{k-2} (2j-1-2is'') \right) \right\}.
\end{eqnarray}
Combining the two, we obtain for $A.B+B.D$:
\begin{eqnarray}
A.B+B.D &=& 0 \quad 2j=2,3,...,k-2  \nonumber \\
    &=&  \frac{1}{2} \delta_{-1-2r,2m''} \delta^+(s'')
\quad 2j=1 \nonumber \\
 &=&   \frac{1}{2} \delta_{-1-2r,2m''} \delta^+(s'')
\quad 2j=k-1.
\end{eqnarray}
Combining this final result with the character identities
in the bulk of the paper, one finds the charge conjugation
matrix.
\boldmath
\subsection*{$S^2$ computation for the bosonic coset}
\unboldmath
Details of the computation of $S^2$ for the coset are as follows. We have :
\begin{eqnarray}
A & = & \frac{i}{\sqrt{k(k-2)}}
e^{ -\frac{ \pi i}{k} (2j+2r) (2j'+2r')} e^{ \frac{ \pi i}{k-2}
  (2j-1)(2j'-1)} \quad for \quad 2j'=2, \dots, k-2 \nonumber \\
 &=&  \frac{i}{2 \sqrt{k(k-2)}} e^{- \frac{\pi i}{k} (2j+2r)(1+2r')}
\quad \mathrm{for} \quad 2j'=1 \dots \nonumber \\
  &=&  -\frac{i}{2 \sqrt{k(k-2)}} e^{- \frac{\pi i}{k} (2j+2r)(-1+2r')}
\quad \mathrm{for} \quad 2j'=k-1 \dots \nonumber \\
D &=&  \frac{2}{\sqrt{k(k-2)}} e^{ \frac{ \pi i}{k} (2 \alpha +2r) (2
  \alpha'+2r')} \cos(\frac{4 \pi s s'}{k-2}) \nonumber \\
B &=&  \frac{1}{\sqrt{k(k-2)}} e^{ \frac{ \pi i}{k} (2j+2r) (2 \alpha'+2r')}
\frac{\cosh  \pi \left( s' \frac{k-4j}{k-2}+i \alpha' \right)}{\cosh \pi (s'+i \alpha')}.
\end{eqnarray}
We can then calculate $A.B$ and $B.D$ respectively, which gives a
matrix labeled by discrete and continuous indices $(j,r)$ and
$(s'',\alpha''+r'')$. We find:
\begin{eqnarray}
A.B &=& \frac{1}{2(k-2)} \delta^{k}_{2j+2r-2 \alpha''-2r''} [p] \nonumber \\
& & 
\left\{ \frac{1+(-1)^p \cos \frac{-i 2 \pi s''+\pi   (1-2j)}{k-2}}{\sin \frac{\pi}{k-2} (1-2j-2is'')}
+\frac{1+(-1)^p \cos \frac{i 2 \pi s''+\pi   (1-2j)}{k-2}}{\sin
    \frac{\pi}{k-2} (1-2j+2is'')} \right\} \nonumber \\
B.D &=&  -\frac{1}{2(k-2)} \delta^{k}_{2j+2r+2 \alpha''+2r''} [p]  \nonumber \\
& &
\left\{ \frac{1+(-1)^p \cos \frac{-i 2 \pi s''+\pi   (1-2j)}{k-2}}{\sin
  \frac{\pi}{k-2} (1-2j-2is'')}
\right.\nonumber\\
&&\left. \qquad \qquad  \qquad +\frac{1+(-1)^p \cos \frac{i 2 \pi s''+\pi   (1-2j)}{k-2}}{\sin
    \frac{\pi}{k-2} (1-2j+2is'')} \right\}
\quad \mathrm{for} \quad 2j=2,\ldots ,k-2 \nonumber \\
& & = \frac{1}{2} \delta^{2k}_{2r+1+2\alpha''+2r''} \delta^+(s'') \quad
\mathrm{for} \quad 2j=1 \nonumber \\
& & = \frac{1}{2} \delta^{2k}_{2r-1+2\alpha''+2r''} \delta^+(s'')
\quad \mathrm{for} \quad 2j=k-1.
\end{eqnarray}
Combining these results with the character identities in the bulk
of our paper yields the
result $S^2=C$, where $C$ is the matrix that exchanges the characters
labeled $(j,r)$ and $(\frac{k}{2}-j,-r)$ in the discrete sector and
$(s,\alpha+r)$ and $(s,-(\alpha+r))$ in the continuous sector.

\end{document}